\documentstyle[prb,aps,multicol,psfig]{revtex}
  \renewcommand{\narrowtext}{\begin{multicols}{2} \global\columnwidth20.5pc}
  \renewcommand{\widetext}{\end{multicols} \global\columnwidth42.5pc}

\newcommand{\f}[1]{Fig.~\ref{#1}}
\newcommand{\eq}[1]{Eq.~(\ref{#1})}
\newcommand{\eqs}[2]{Eqs.~(\ref{#1}) and~(\ref{#2})}

\def\be{\begin{equation}}
\def\ee{\end{equation}}
\def\bea{\begin{eqnarray}}
\def\eea{\end{eqnarray}}
\def\l({\left(}
\def\r){\right)}
\def\p{\protect}
\def\Bn{B_0}
\def\Bam{B_{am}}
\def\Jc2{\tilde J_c}
\def\Bceff{B_c^{\text{eff}}}

\begin{document}
\bibliographystyle{prsty}
\draft

\title {Thin superconducting disk with
$B$-dependent $J_c$ : Flux and current 
distributions}
\author{D.~V. Shantsev$^{1,2}$, Y.~M.~Galperin$^{1,2}$, and T.~H.~Johansen$^1$
}

\address{$^1$Department of Physics, University of Oslo, P. O. Box 1048 Blindern, 
0316 Oslo, Norway\\
$^2$A. F. Ioffe Physico-Technical Institute, Polytechnicheskaya 26, 
St.Petersburg 194021, Russia}

\date{\today}

\maketitle

\begin{abstract}
The critical state in a superconducting thin circular disk with an arbitrary
magnetic field dependence of the critical sheet current, $J_c(B)$, is analyzed.
With an applied field $B_a$ perpendicular to the disk, a set of coupled integral 
equations for the flux and current distributions is derived. The equations are 
solved numerically, and flux and current profiles are presented graphically
for several commonly used $J_c(B)$ dependences. 
It is shown that for small $B_a$ the flux penetration depth
can be described by an effective Bean model with a renormalized $J_c$ entering
the leading term.
We argue that these results are qualitatively correct for {\em thin} 
superconductors 
of {\em any shape}. The results contrast the parallel geometry behavior,
where at small $B_a$ the $B$-dependence of the critical current can be ignored.

\end{abstract}


\narrowtext

\section{Introduction} 

The critical state model (CSM) is widely accepted to be a
powerful tool in the analysis of magnetic
properties of type-II superconductors. 
For decades there have been numerous theoretical works
devoted to CSM calculations in the 
parallel geometry, i. e., a long sample placed
in parallel applied magnetic field, $B_a$. 
More recently, much attention has also been paid
to the CSM analysis of thin samples
in perpendicular magnetic fields. For this so-called perpendicular geometry
explicit analytical expressions for flux and current distributions
have been obtained for a
long thin strip~\cite{BrIn,Zeld} and thin circular 
disk\cite{Mik,Zhu,Clem,MikNote} assuming 
a constant critical current (the Bean model). 

{}From experiments, however, it is well known that the critical current
density $j_c$ usually depends strongly on the local flux density $B$. 
This dependence often hinders a precise interpretation of various 
measured quantities such as magnetization, complex ac 
susceptibility,\cite{Gomory,Rao} and surface impedance.\cite{Willemsen}
It is therefore essential to extend the CSM analysis to account for a
$B$-dependence of $j_c$. 

In the parallel geometry extensive work has already been carried
out, and exact results for the flux density profiles and 
magnetization,\cite{Clem79,Forsthuber,Xu,JohBra}, as well as 
ac losses\cite{Clem79,Forsthuber} have been obtained for
different $j_c(B)$-dependences.
In the perpendicular geometry the magnetic behavior 
is known to be qualitatively different.
In particular, due to a strong demagnetization, the field 
tends to diverge at the sample edges, and
the flux penetration depth and ac losses 
follow different power laws in $B_a$ for small $B_a$.\cite{BrIn,Zeld}
Unfortunately, the theoretical treatment of the perpendicular geometry
is very complicated, and we are not aware of any explicit
expressions obtained for the CSM with a $B$-dependent $j_c$.
However, it is possible to derive integral equations relating the 
flux and current distributions.\cite{Mik} Such equations have 
so far been obtained and solved numerically only for the case of a 
long thin strip.\cite{McD,1}

In this paper, we derive a CSM solution for a thin circular
disk characterized by an {\em arbitrary\/} $j_c(B)$.  
The solution is presented as a set of integral equations which we 
solve numerically. 
In this way we obtain the field and current density distributions 
in various magnetized states. We present results for several commonly
used functions $j_c(B)$. A special attention is paid
to the low-field asymptotic behavior.

The paper is organized as follows. In Sec.~\ref{be} the basic
equations for the disk problem are
derived. We consider here all states during a complete cycle of the
applied field, including 
the virgin branch. Sec.~\ref{fcd} contains our numerical results for flux and
current distributions as well as for the flux front position. 
A discussion of the results is presented. 
Finally, Sec.~IV presents the conclusions. 

\section{Basic equations }
\label{be}

Consider a thin superconducting disk of radius $R$ and thickness $d$, 
where $d \ll R$, see \f{f_disk}. We assume either that $d \ge \lambda$,
where $\lambda$ is the London penetration depth, or, if $d < \lambda$,
that $\lambda^2/d \ll R$. In the latter case the quantity $\lambda^2/d$ 
plays a role of two-dimensional penetration depth.~\cite{DeGennes} 
We put the origin of the reference frame at the disk center and direct 
the $z$-axis perpendicularly to the disk plane.  
The external magnetic field ${\bf B}_a$ is applied along the $z$-axis,
the $z$-component of the field in the plane $z=0$ being denoted
as $B$. The current flows in the azimuthal direction, with a
sheet current denoted as $J(r)=\int_{-d/2}^{d/2} j(r,z)\, dz$, where
$j$ is the current density.

To obtain expressions for the current and flux distribution we follow
a procedure originally suggested in Ref.~\onlinecite{Mik} and then
generalized in Ref.~\onlinecite{McD} for the case of $B$-dependent
$J_c$ in a thin strip. 
The procedure makes use of the  Meissner state distributions for $B$ and $J$.

In the Meissner state, where $B=0$ inside the disk, 
the field outside the disk is given by~\cite{Mik,Zhu} 
\be
  B_M(r,R) = B_a + \frac{2B_a}{\pi} \left[ 
\frac{R}{\sqrt{r^2-R^2}} -
\arcsin\left(\frac{R}{r}\right) 
   \right] ,
\label{BM}
\ee
and the current is distributed according to
\be
  J_M(r,R) = - \frac{4 B_a}{\pi \, \mu_0} \frac{r}{\sqrt{R^2-r^2}}, \quad r<R \ 
.
\label{JM}
\ee

\input{fig0.inp}
 
\subsection{Increasing field \label{vs}}

We begin with a situation where
the external field $B_a$ is applied to a zero-field-cooled disk.
The disk then consists of an inner  flux-free region, $ r \le a$ ,
and of an outer region, $a < r \le R$, penetrated by magnetic flux. According to 
the
critical state model, the penetrated part will carry the 
critical sheet current $J_c$ corresponding to the {\em local\/} value of
magnetic field,   
\be
 J(r) = -J_c[B(r)], \quad  a<r<R
\label{J0}
\ee
Now following Refs.~\onlinecite{Mik,McD}, we express the field and current as 
superpositions of the Meissner-state distributions, (\ref{BM}) and
(\ref{JM}), i.e.: 
\bea
  B(r)& =& \int^{\min(r,R)}_a \! dr' \, B_M(r,r')\,
    G(r', B_a)\, . \label{BG} \\
J(r)& =& \int^R_{\max(a,r)} \! dr' \, J_M(r,r')\, G(r', B_a)\, ,
\label{JG} 
\eea
where $G(r,B_a)$ is a weight function.
Since $B(r)$ and $B_M (r,R) \rightarrow B_a$ at $r \rightarrow \infty$ we have
 the normalization condition 
\be
\int_a^R \! dr \, G(r,B_a) =1\, . \label{nc}
\ee
Substituting \eq{J0} and \eq{JM} into~\eq{JG} yields an integral equation for 
the 
function $G(r,B_a)$ which can be inverted to obtain 
\be \label{G}
 G(r, B_a) = - \frac{B_c}{B_a} \frac{d}{dr} \int_r^R
   \frac{dr'}{\sqrt{r'^2 - r^2}}
   \frac{J_c[B(r')]}{J_{c0}},
\ee
where 
\be
  B_c = \mu_0 J_{c0} /2\, , \quad J_{c0} \equiv J_c(B=0).
\ee
Note that due to the similar form of the function $J_M(r,R)$ in (\ref{JM})
and the Meissner-state current in the strip case\cite{BrIn,Zeld} 
our weight function $G(r, B_a)$  is also similar to that for a strip, see Ref.~\onlinecite{McD}

{}From \eqs{JG}{G} it then follows that the 
current distribution in a disk is given by
\be
  J(r) = \left\{
  \begin{array}{lr}
  - \displaystyle{
    \frac {2r}{\pi} 
         \int_a^R\! \!
 dr' 
\sqrt{\frac{a^2 - r^2}{r'^2 -a^2}}\,\frac{J_c[B(r')]} {r'^2 - r^2}}
 , & r<a\\
         - J_c[B(r)], &\! \! a<r<R
  \end{array}
  \right.
\label{JV}
\ee
This equation is supplemented by
the Biot-Savart law, which for a disk reads,\cite{Mik}
\be
  B(r) = B_a + \frac{\mu_0 }{2\pi} \int^R_0 F(r,r') J(r')  dr'\,.
\label{BJ}
\ee
Here $F(r,r') = K(k)/(r+r') -
E(k)/(r-r')$,  where $k(r,r') = 2\sqrt{rr'}/(r+r')$, while $K$ and 
$E$ are complete elliptic integrals
defined as $E(z)=\int_0^{\pi/2} \! \left[ 1-z^2 \cos^2(x)\right]^{1/2} dx$ and 
$K(z)=\int_0^{\pi/2}\! \left[ 1-z^2 \cos^2(x)\right] ^{-1/2} dx$.

The relation between the flux front location $a$ and applied field
$B_a$ is obtained by substituting \eq{G} into \eq{nc}, giving
\be
   B_a = B_c \int_a^R \frac{dr'}{\sqrt{r'^2-a^2}} \frac{J_c[B(r')]}{J_{c0}}.
\label{Ba}
\ee

For a given $B_a$ and for a specified $J_c(B)$  we need to solve the 
set of three coupled equations (\ref{JV})-(\ref{Ba}).
In the case of $B$-independent $J_c$, the \eq{Ba} acquires the simple form
\begin{equation} \label{a}
a/R=1/\cosh (B_a/B_c)\, ,
\end{equation} 
and the \eqs{JV}{BJ} lead to the
Bean-model results derived in  Refs.~\onlinecite{Zhu}
and~\onlinecite{Clem}.  
 
Note that the equations  can be significantly simplified at
large external field where  $a \rightarrow 0$ proportionally  to
$\exp(-B_a/B_c)$. Then
$B(r)$ is determined by the single equation 
\begin{equation} \label{sm_a}
B(r)=B_a -\frac{\mu_0 }{2\pi} \int^R_0 F(r,r') J_c[B(r')]  dr'\, ,
\end{equation}
following from \eq{BJ}.

\subsection{Subsequent field descent}

Consider now the behavior of the disk as $B_a$ is
reduced after being first raised to some maximal value $\Bam$.
Let us denote the flux front position, the current density and the
field distribution  at the maximum field as $a_m$, $J_m(r)$
and $B_m(r)$, respectively. 
Obviously, $J_m(r)$, $B_m(r)$, and $a_m$ satisfy
Eqs. (\ref{JV})-(\ref{Ba}).

\input{fig01.inp}

During the field descent from $\Bam$ the flux density becomes reduced 
in the outer annular region $a<r<R$, see \f{f_scheme}. The central 
part of the disk $r<a$ remains frozen in the state with $B_a =\Bam$. 
Let us specify the field and current distributions in this remagnetized
state as 
\be
 B(r) = B_m(r) + {\tilde B}(r), \quad J(r) = J_m(r) + {\tilde J}(r),
\label{01}
\ee
and derive the relation between ${\tilde B}(r)$ and ${\tilde
J}(r)$. For that one 
can use a procedure similar to the one described in Sec.~\ref{vs}.
The only difference is that in the region $a<r<R$~ we now have to use
 $J(r) = + J_c[B(r)]$. In this way we obtain
$$
{\tilde J}(r) = \Jc2(r), \quad \quad \quad a<r<R,
$$
where we define
\be
\Jc2(r) = J_c[B_m(r)+{\tilde B}(r)] + J_c[B_m(r)]\,. \label{Jeff}
\ee
Note that the function $\Jc2(r)$ depends on the coordinate
only  through the field distributions $B_m(r)$ and ${\tilde B}(r)$.
Instead of \eq{JV}, the additional current satisfies
\be
  {\tilde J}(r) = \left\{\!
  \begin{array}{lr}
  \displaystyle{
  \frac {2r}{\pi}  
         \int_a^R \!dr' \ \sqrt{\frac{a^2 - r^2}{r'^2 - a^2}}\,
  \frac{\Jc2 
  (r') } {r'^2 - r^2}} ,
         & r<a\\
         \Jc2(r), & a<r<R
  \end{array}
  \right.
\label{JR}
\ee
with the complementary equation 
\be
  {\tilde B}(r) = B_a-\Bam + \frac{\mu_0 }{2\pi} \int^R_0 F(r,r') {\tilde J}(r')  
dr' \, .
\label{B1J}
\ee
Furthermore, similarly to \eq{Ba}, we have
\be
   \frac{B_a - \Bam}{B_c} = -  \int_a^R\!  \frac{dr}{\sqrt{r^2-a^2}} \
               \frac{\Jc2 (r)  }{J_{c0}} \ ,
\label{Ba1}
\ee
which completes the set of equations describing the remagnetized state.
Again, for $B$-independent $J_c$  the equations reproduce the Bean-model
results~\cite{Zhu,Clem} 

If the field is decreased below $-\Bam$ the memory of 
the state at $B_a =\Bam$ is completely 
erased, and the solution becomes equivalent to the virgin penetration case.
If the difference
$\Bam-B_a$ is sufficiently large then $a \rightarrow 0$ rapidly, and the 
critical state $J(r)=J_c(r)$ is established throughout the disk. 
In this case the field descent is described by \eq{sm_a} with the 
opposite sign in front of the integral.  

We emphasize that the expressions derived here (Sec.II A and B)
are readily converted to the long thin strip case. This is due to 
the similarity of (\ref{JM}) and the expression for the Meissner-current
in a strip, 
\be
  J(x) = - \frac{2 B_a}{\mu_0 } \frac{x}{\sqrt{w^2-x^2}}, \quad -w<x<w,
\label{JMstrip}
\ee
where $x$ is the coordinate across the strip.
Thus, making in this paper the substitutions
$$r \rightarrow x, \ R \rightarrow w, \ F(r,r') \rightarrow
\frac{2x'}{x'^2 -x^2}, \ B_c \rightarrow \frac{\mu_0 J_{c0}}{\pi}$$ 
one immediately arrives at the set of equations valid for a thin 
long strip. In that case some of 
the integrals can be done analytically to yield the expressions
obtained in Ref.~\onlinecite{McD}.  

A difference between the
derivation in Ref.~\onlinecite{McD} and the present one is that for decreasing 
fields
we calculate only the additional field ${\tilde B}(r)$ rather
than the total field $B(r)$. This
allows us to use only one weight function (\ref{G}) to calculate the  
flux distributions both for increasing and decreasing fields.  This simplifies
the numerical calculations significantly.  

\subsection{Numerical procedure}

Given the $J_c(B)$-dependence, the magnetic behavior
is found by solving the derived integral equations (\ref{JV})-(\ref{Ba}) 
numerically using the following iteration procedure.
With $B_a$ increasing a flux front position $a$ is first
specified, and an initial approximation  for $B(r)$, e.g., the Bean model
solution, is chosen.
At each step the $n$th approximation, $B^{(n)}(r)$, is used to calculate
$J^{(n)}(r)$ from \eq{JV} and $B_a^{(n)}$ from \eq{Ba}.
They are  then substituted into \eq{BJ}
yielding the next approximation, $B^{(n+1)}(r)$.
The iterations are stopped when $\left(R^{-1}\int
dr\,\left[B^{(n+1)}(r) - B^{(n)}(r)\right]^2 \right)^{1/2}\le 10^{-6}B_c$.
With $B_a$ decreasing, the same procedure is used to find first
$J_m(r)$, $B_m(r)$, $\Bam$ for a given $a_m$.
Then, Eqs.~(\ref{JR})-(\ref{Ba1}) are solved for a fixed
$a$ yielding the functions ${\tilde B}(r)$ and ${\tilde J}(r)$ and also
the applied field $B_a$.

\section{Flux and current distribution }
\label{fcd}

\subsection{General features \label{gf}}
In the numerical calculations we used
the following dependences $J_c(B)$,
%
\begin{figure} 
\vbox{
\centerline{ \psfig{figure=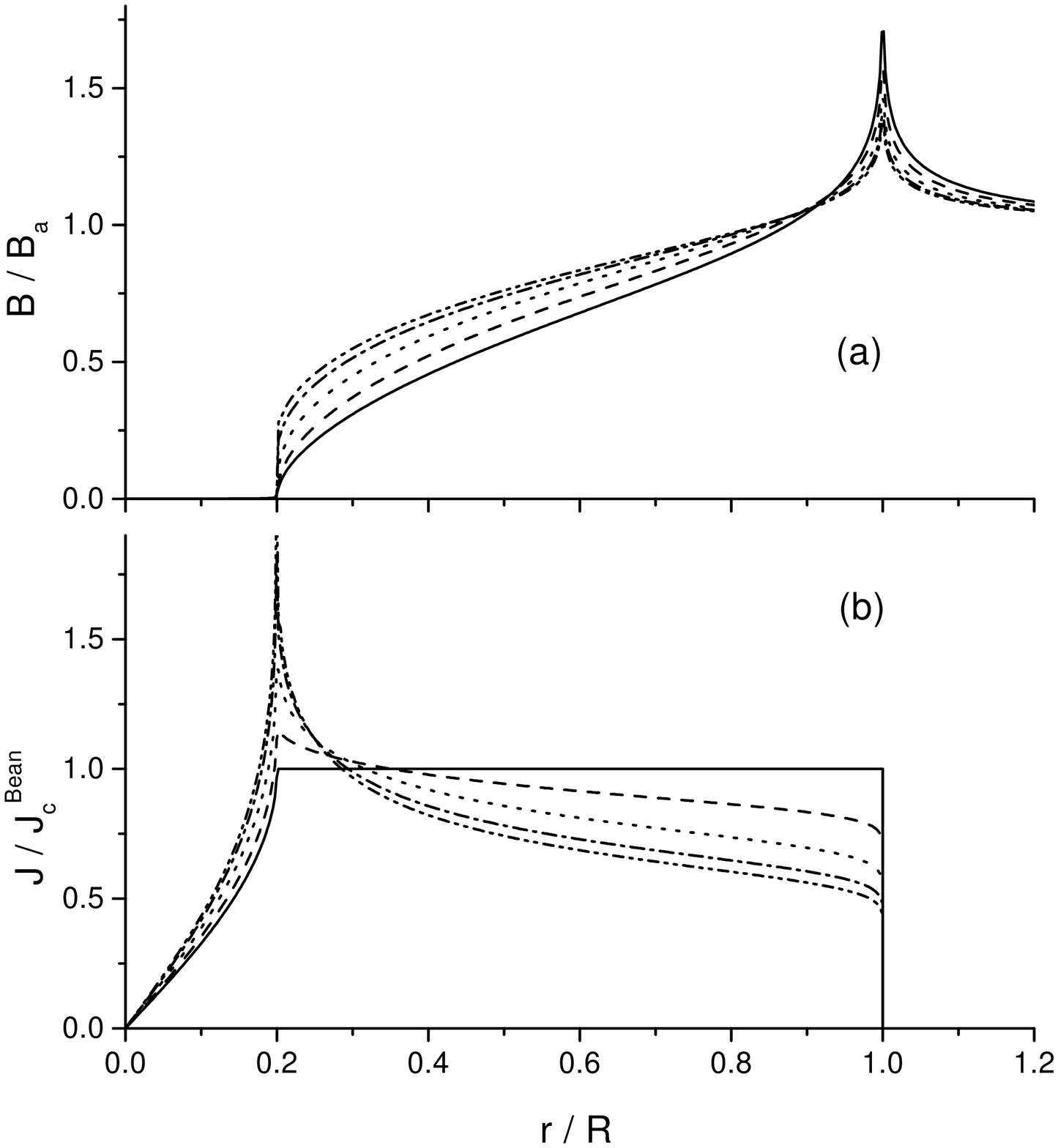,width=8cm}}
\centerline{ \psfig{figure=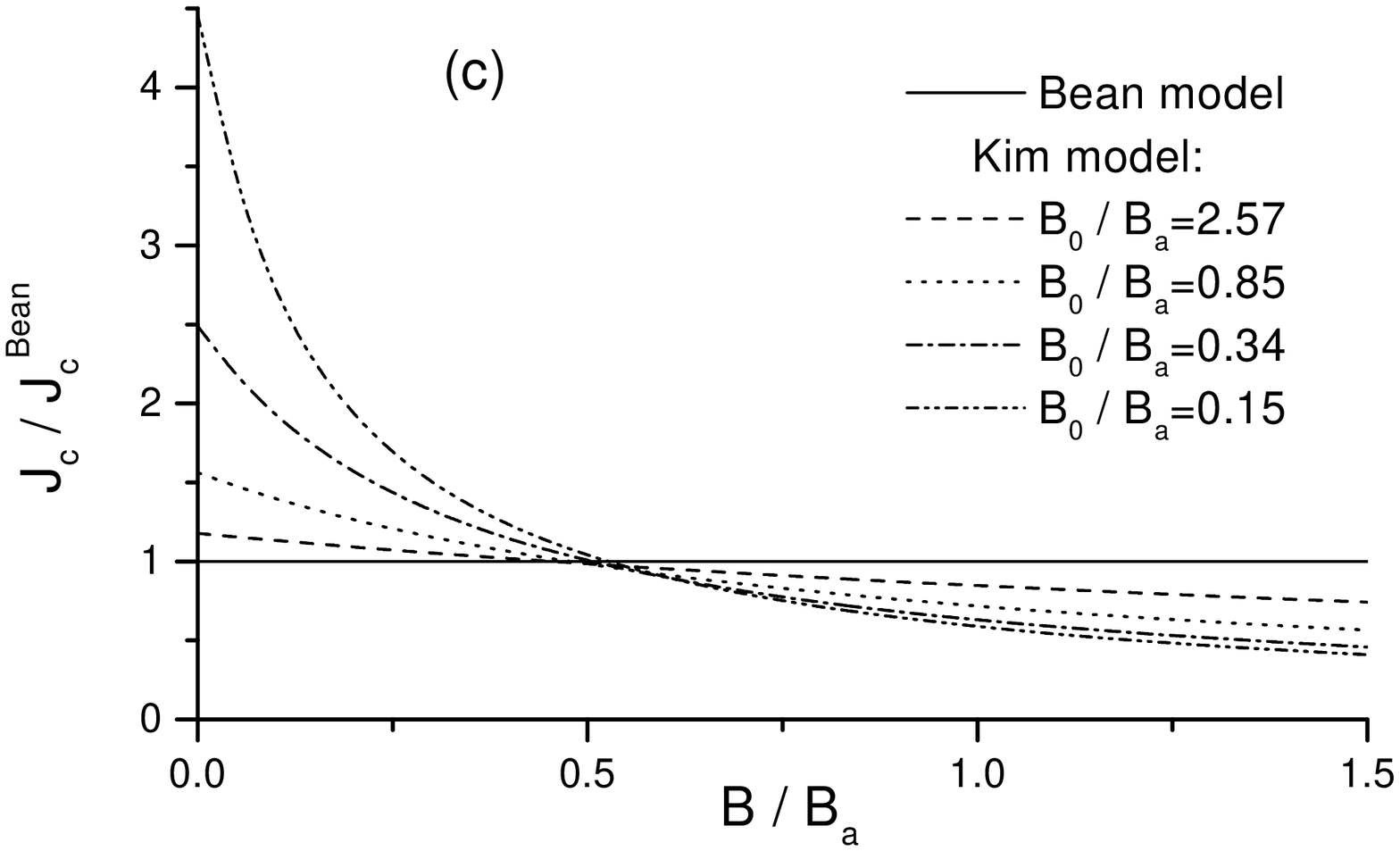,width=8cm}}
\caption{Flux density (a) and current (b) profiles for flux
penetration into a virgin state. Solid lines are calculated for the Bean
model, while other ones represent the Kim model, \p{\eq{km}},
with different $J_{c0}$ and $\Bn$. These parameters are chosen so that
the flux front position $a=0.2 R$ and applied field $B_a$ are the 
same for all the profiles.
The corresponding   $J_c(B)$-dependences are shown in the panel (c). 
The current is normalized to the Bean 
model critical current $J_c^{\rm Bean}$. \label{f_bkim}}
}
\end{figure}

\bea
J_c&=& J_{c0}/(1+|B|/\Bn) \quad \quad \ \text{(Kim model),} \label{km} \\
J_c&=& J_{c0}\, \exp(- |B|/\Bn) \qquad   \text{(exponential model).}
\label{em}
\eea

Shown in Fig.~\ref{f_bkim} (a,b) are
the field and current distributions for increasing field
with $a/R=0.2$
for the Kim model with different parameters $J_{c0}$ and $\Bn$. They
are chosen in a way to keep the position $a$ fixed for
all the
curves for a given value of $B_a$. This allows us to follow
the variations in the profile shape as the
$B$-dependence of $J_c$ changes.
The chosen parameters   $J_{c0}$ and $\Bn$ correspond to
the set of $J_c(B)$-curves shown in  Fig.~\ref{f_bkim} (c).  
The Bean model results are also plotted in \f{f_bkim}.
\input{fig2.inp}

Several major deviations between the Kim and the
Bean model can be noticed. In the Kim model we see that
(i) the current $J(r)$ is not uniform at $a<r<R$ --
it is minimal at the disk edge where $|B|$ is maximal;
(ii)  the current has a cusp-like maximum at $r=a$ since the
magnetic field vanishes at this point with infinite derivative;
(iii) compared to the Bean model,
the $B(r)$ profiles are steeper near the flux front, whereas
the peaks at the edges are less sharp.
\input{fig3.inp}
Qualitatively similar results are obtained for the exponential model, 
see \f{f_be}. Also here, by changing the model parameters 
one can produce a variety of flux and current profiles which are quite
different from the Bean model predictions. 
When comparing the Kim and exponential model, however,
it turns out to be very difficult to find clear distinctions.

During field descent the $B$- and $J$-profiles become more complicated.
For brevity we show only profiles for the Kim model with $\Bn/B_c=3$
and for the Bean model at different values of $B_a$, see \f{f_br(r)}.
Again, the Kim model gives a non-uniform current density at
$r>a$. Contrary to the increasing-field states, the
current density can now either decrease, or increase towards the edge
depending on $B_a$.

\begin{figure} 
\vbox{
\centerline{ \psfig{figure=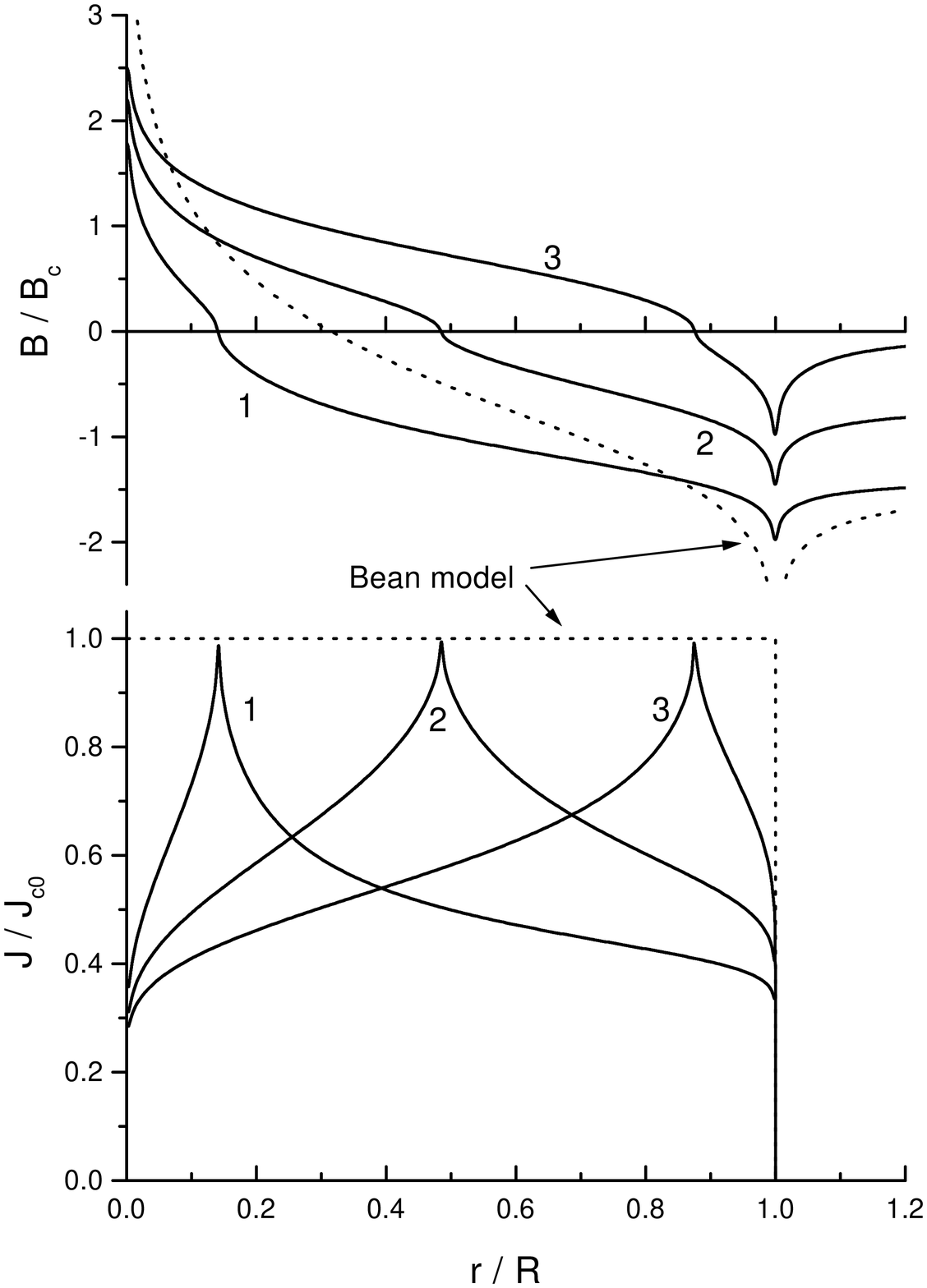,width=7.6cm}}
\caption{Flux and current density profiles
during field descent after first increasing 
$B_a$ to a very large $\Bam$. It is assumed that critical state
is established throughout the disk. Solid lines show results
for the Kim model with $\Bn=B_c$ and $B_a/B_c=-1.4$ (curve 1),
$B_a/B_c=-0.7$ (curve 2), and $B_a = 0$ (curve 3). The dotted lines show the
Bean model profiles, their shape being independent of $B_a$.
In contrast, the shape of the Kim model profiles depends strongly on $B_a$.
 \label{f_fp}}
}
\end{figure}

Figure~\ref{f_fp} shows profiles for fully-penetrated decreasing-field state.
In the Bean model the current remains constant, while
the profile of the flux distribution is fixed, although
shifted according to the applied field.
In contrast, the Kim-model profiles are strongly dependent on $B_a$.
There is a peak in the current profile and an enhanced gradient of
$B(r)$  near the point where $B=0$.

\subsection{Flux penetration depth \label{fpd}}

To analyze quantitatively the role of a $B$-dependent $J_c$
let us consider the position of the flux front during increasing
field. A circle with radius $a$ then limits the Meissner region $B=0$,
and is also the location of maximum gradient in $B$.
These features can be measured directly 
in experiments on visualization of magnetic flux distribution,
e. g., magneto-optical imaging.\cite{MO} 

Let us first recall the CSM expression for the flux
front location in a long
circular cylinder in a parallel field~\cite{JohBra},
\be \label{cyl1}
\frac a R =1 - \frac{1}{\mu_0 R}\int_0^{B_a} \! \frac{dB}{j_c(B)} \, .
\ee
At small applied fields, $B_a$, it can be expanded as
\be \label{cyl2}
\frac a R \approx 1-\frac{B_a}{B_c}+ \frac{\mu_0 R}{2}  \, j_c'(0) \,
\left(\frac{B_a}{B_c}\right)^2\,,    
\ee 
where for a cylinder $B_c \equiv \mu_0 j_{c0} R$. 
Note that the $B$-dependence of $j_c$ enters the expansion
first in the second-order term.
Consequently, for a long cylinder the low-field behavior of $a$ is well 
described by the Bean model where the penetration depth increases linearly with 
the applied field.

For a thin disk, the penetration  of flux proceeds differently.
In the Bean model the location of the flux front, \eq{a}, is for
small $B_a$ given by
\be \label{a_s}
a/R \approx 1- (1/2)\, (B_a/B_c)^2\, .
\ee
For an {\em arbitrary} $B$-dependence of  $J_c$
the expression (\ref{Ba}) relating $a$ and $B_a$
cannot easily be expanded in powers of the ratio $B_a/B_c$.
The physical reason for this is the singular behavior of the
magnetic field near the disk edge. There, the local field diverges
at {\em any\/} finite $B_a$, and an expansion of $J_c(B)$ in
powers of $B$ is not everywhere convergent. To clarify the behavior of
the flux front we have therefore performed
numerical calculations of the dependences $a(B_a,\Bn)$. Shown in \f{f_3d}
are the results for the  Kim (upper panel) and the
exponential (lower panel) models.
Note that the limit of large $\Bn$ represents the Bean model.

For small $B_a$ all the models seem to yield a parabolic relation
between the penetration depth and the applied field.
This is illustrated in more detail in
\f{f_slope2}, where all the graphs in the log-log plot have a slope of 2
in the low-field region.
We therefore conjecture that {\em any} $B$-dependence of $J_c$ leads
to the {\em same\/} quadratic law (\ref{a_s}) as for the Bean
model, although with {\em different\/} coefficients in front of $(B_a/B_c)^2$.
\begin{figure} 
\vbox{
\centerline{ \psfig{figure=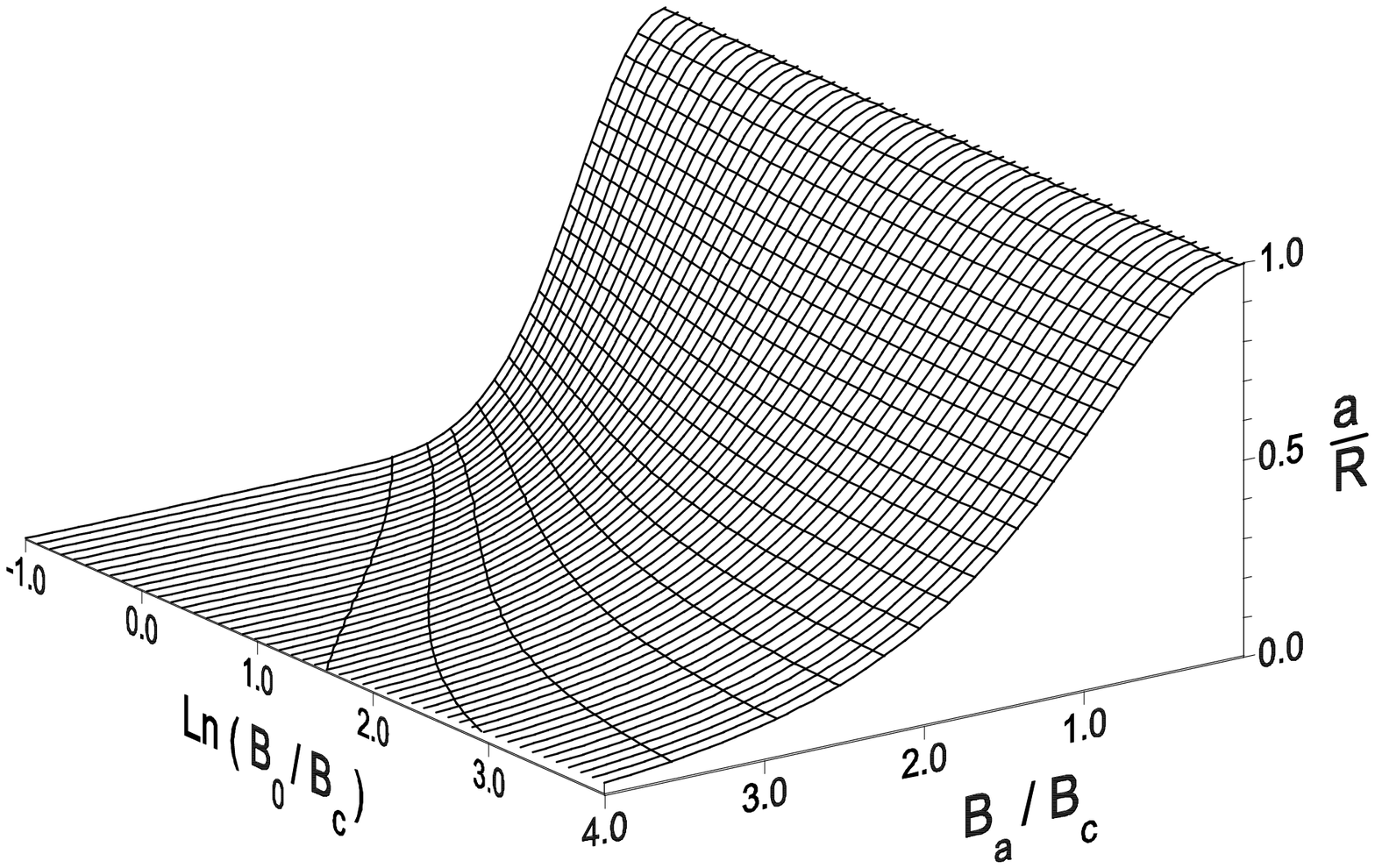,width=8cm}}
\centerline{ \psfig{figure=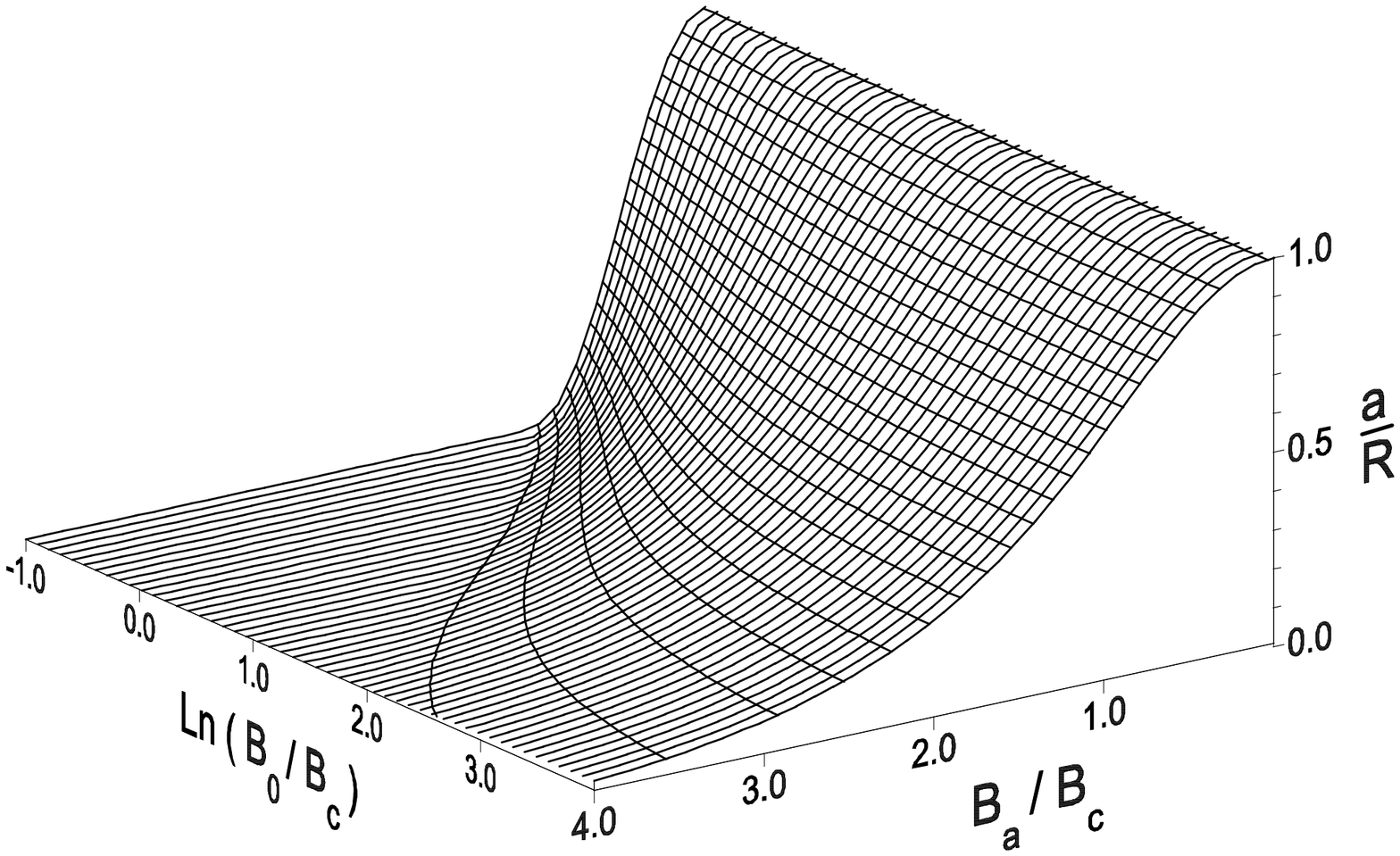,width=8cm}}
\caption{\label{f_3d}
Flux front location $a$ as a function of increasing applied field, $B_a$,
and the charcteristic field, $\Bn$. The calculated results are presented for the
Kim model (upper panel), \eq{km}, and for the exponential model (lower panel),
\eq{em}.}
}
\end{figure}
\input{fig6.inp}

The overall behavior of the penetration depth can be fitted well
by the full form
\eq{a}, provided one makes the substitution $B_c \rightarrow \Bceff$, i.e.,
\be \label{if}
a/R = 1/ \cosh (B_a/\Bceff ) \ .
\ee
We find that the effective $B_c$ satisfies the relation
\be \label{if2}
\frac{\Bceff}{B_c} = 1-\alpha \, \frac{ B_c}{  \Bn} \, ,
\ee
if the ratio $B_0/B_c$ is of the order 1, or larger.
Here $\alpha = 0.42$ for the exponential model, and $\alpha = 0.36$
for the Kim model.
The same relations~(\ref{if}) and~(\ref{if2}) are found to hold true for
a long thin strip with $\alpha = 0.60$ and 0.51 for the
exponential and the Kim model, respectively.

We believe that for many purposes a Bean-model description with an effective
critical current is appropriate for thin samples of any shape,
both in applied field and under transport current.
Indeed, strong demagnetization effects always lead to a
divergence of magnetic field at the sample edge.
This implies that in the sample there is always present
a wide range of $B$ values up to infinity. As a result, the sample behavior
is determined by the whole $J_c(B)$-dependence.
In particular, the value $J_c(0)$ is not governing the magnetic behavior
of thin samples, even when the applied field is very small.\cite{d}

\section{Conclusion}

A set of integral equations for the magnetic flux and current
distributions in a thin disk placed in a perpendicular applied field
is derived within the critical state model. The solution is valid
for any field-dependent critical current, $J_c(B)$.
By solving these equations numerically it is demonstrated that
both the flux density and current profiles are
sensitive to the $J_c(B)$-dependence.
In particular, compared to the Bean model,
the $B(r)$-profiles are steeper near the flux front, whereas
the peaks at the edges are less sharp.

Since the local magnetic field at the disk edge is divergent for any
value of the applied field, $B_a$, a field dependence of $J_c$ affects
the flux distribution {\em even in the limit of low} $B_a$.
Our numerical calculations show that the flux penetration depth at
small fields has the same quadratic dependence on $B_a$ as for the Bean
model, however with different coefficient.
The overall behavior of the flux penetration depth is
well described by the Bean model expression with an effective value of the
critical current.
These results are believed to be qualitatively correct for {\em thin}
superconductors of {\em any shape}. The behavior differs strongly from
the case of a long cylinder in a parallel field, where the front position
at low $B_a$ is not affected
by the $B$-dependence of the critical current density.

\acknowledgements

The financial support from the Research Council of Norway
is gratefully acknowledged.


\widetext 
\begin{references}
\bibitem{MikNote} Some final expressions obtained in Ref.~\onlinecite{Mik}
are not correct, see discussion in Ref.~\onlinecite{Zhu}.
\bibitem{1}
An alternative approach to treat the problem is to carry out flux
creep simulations.\cite{Br-ring,Br-disk1}  Here the CSM results can be found as 
a limiting case of a large creep exponent.
\bibitem{d}
In samples of finite thickness $d$ the field divergency at the sample
edge is cut off at values of the order of $B_c \ln (w/d)$.
Thus, the low-field behavior is determined by 
$J_c(B)$-dependence at fields less than this value.
\end{references}
\end{document}